\documentclass[usenatbib,hyperref]{mnras}
\usepackage{multirow}
\usepackage[T1]{fontenc}
\usepackage{ae,aecompl}
\usepackage{amsmath}
\usepackage{etoolbox}
\usepackage{breqn}
\usepackage{graphicx}
\usepackage{textcomp}
\usepackage{subfigure}
\usepackage{adjustbox}
\usepackage[flushleft]{threeparttable}

\newcommand{\mr}{\mathrm}

\renewcommand{\degr}{$^\circ$}
\newcommand\arcs{\ensuremath{^{\prime\prime}}}

\newcommand{\fermi}{\emph{Fermi}-LAT}
\newcommand{\hess}{H.E.S.S.}

\newcommand\cutoff{cutoff}

\title[Prospects for the characterization of the VHE emission from the Crab Nebula with the CTA]{Prospects for the characterization of the VHE emission from the Crab Nebula and Pulsar with the Cherenkov Telescope Array}
\author[E. Mestre et al.]{E.~Mestre$^{1,2}$,
E.~de O\~na Wilhelmi$^{1,2,3}$,
R.~Zanin$^{4}$, D. F. Torres$^{1,2,5}$
and  L. Tibaldo$^{6}$\\
$^{1}$Institute of Space Sciences (ICE/CSIC), Campus UAB, Carrer de Can Magrans s/n, 08193 Barcelona, Spain\\
$^{2}$Institut d' Estudis Espacials de Catalunya (IEEC), 08034 Barcelona, Spain\\
$^{3}$Deutsches Elektronen Synchrotron DESY, 15738 Zeuthen, Germany\\
$^{4}$Max-Planck-Institut f\"ur Kernphysik, P.O. Box 103980, D 69029 Heidelberg, Germany\\
$^{5}$Instituci\'o Catalana de Recerca i Estudis Avan\c{c}ats (ICREA), E-08010 Barcelona, Spain\\
$^{6}$Institut de Recherche en Astrophysique et Plan\'etologie, CNRS-INSU, Universit\'e Paul
}

\begin{document}

\maketitle
\date{Received 2019 November 27; in original form 2019 August 21 ; accepted 2019 November 29}
\pubyear{2018}

\begin{abstract}
The Cherenkov Telescope Array (CTA) will be the next generation instrument for the very high energy gamma-ray astrophysics domain. With its enhanced sensitivity in comparison with the current facilities, CTA is expected to shed light on a varied population of sources.
In particular, we will achieve a deeper knowledge of the Crab nebula and pulsar, which are the best characterized pulsar wind nebula and rotation powered pulsar, respectively. We aim at studying the capabilities of CTA regarding these objects through simulations, using the main tools currently in development for the CTA future data analysis: {\it Gammapy} and {\it ctools}.
We conclude that, even using conservative Instrument Response Functions, CTA will be able to resolve many uncertainties regarding the spectrum and morphology of the pulsar and its nebula. 
The large energy range covered by CTA will allow us to disentangle the nebula spectral shape among different hypotheses, corresponding to different underlying emitting mechanisms. In addition, resolving internal structures (smaller than $\sim$ 0.02\degr{} in size) in the nebula and unveiling their location, would provide crucial information about the propagation of particles in the magnetized medium. 
We used a theoretical asymmetric model to characterise the morphology of the nebula and we showed that if predictions of such morphology exist, for instance as a result of hydrodynamical or magneto-hydrodynamical simulations, it can be directly compared with CTA results. We also tested the capability of CTA to detect periodic radiation from the Crab pulsar obtaining a precise measurement of different light curves shapes.

\begin{scriptsize}\begin{tiny}\end{tiny}\end{scriptsize}
\end{abstract}

\section{Introduction}

Due to the high luminosity and seemingly long-term flux stability, the Crab pulsar and its pulsar wind nebula (PWN) is one of the most studied sources in the very high energy (VHE, $\rm{E} > 100$ GeV) regime.
For many years, Crab has been used as a standard candle in X- and gamma-ray astronomy (\citealt{Hester2008}). 
The Crab nebula was the first TeV gamma-ray source discovered (in $1989$ by the Whipple $10$ meters telescope, \citealt{1989Weekes}),
and soon after detected by numerous facilities above 100\,GeV \citep{2000Smith,Aharonian2006,MILAGRO_CRAB,Aleksic2015,2015Meagher,HAWC_CRAB}.
It has a (energy-dependent) angular size of $\sim$ 0.1\degr{} and its distance has been estimated to be $\approx 2.2$ kpc, corresponding to a physical size of $\approx 3.8$ pc \citep{1973Trimble,1985Davidson,2008Kaplan}.
The nebula non-thermal spectrum can be described by two components, a synchrotron component extending from radio to high energy gamma-rays and a second component emerging above 1\,GeV (\citealt{1996Atoyan}).
The latter is interpreted as inverse Compton scattering (IC) of the same particles against soft background photons: cosmic microwave background (CMB), far-infrared (FIR) and near-infrared (NIR) background, and the synchrotron photons of the nebula itself or Synchrotron Self Compton (SSC). 
The pulsar has a spin period of $\rm{P} = 33$ ms, a spin down rate of $\rm{\dot{P}} = 4.21\times10^{-13}$ and a spin down luminosity of $\rm{L_{spin}} = 3.8\times10^{38}\ \rm{erg}\ \rm{s^{-1}}$.
The pulsed emission between 0.1\,GeV and 100\,GeV is believed to be due to synchrotron-curvature radiation (\citealt{2010Abdo, 2016Ansoldi}) and its spectrum is well parametrized by a power law with a sub-exponential cutoff function of spectral index $\gamma_{\rm{P}} = 1.59$, the break located at an energy of about 500\,MeV and curvature index of $\kappa = 0.43$. 
In addition, a power law component emerges above the cutoff extending above 100\,GeV (\citealt{2011ScienceVeritas,2012Aleksic,2016Ansoldi}). 

Despite the high-precision spectral measurements performed by the last generation of IACTs, with points with a 5\% statistical uncertainty at energies below 100\,GeV \citep{Aleksic2015}, the spectral shape of the IC component of the Crab nebula is still not firmly established. The main ambiguities appear at the highest energies, above tens of TeV. They are the result of the decrease of the photon statistics, and, on the other hand, of the large systematic uncertainties that the imaging Cherenkov technique suffers. \\
At the lowest energies (from few up to hundreds of GeV), instead, \fermi{} and MAGIC closed the previously existing gap between space- and ground-based measurements, hence providing for the first time a complete coverage of the IC peak. A joint analysis of \fermi{} and MAGIC shows a rather flat peak. Given the small uncertainties, this peak is not trivial to reproduce (see \citealt{Aleksic2015} for an in-depth discussion). Different theoretical models were used to reproduce the rich Crab data sample, following different prescriptions of the time evolution of the system and particle populations (see, i.e. \citealt{1973ApJ...186..249P,1998Hillas,Aharonian2004,2010Meyer,2010Tanaka,Bucciantini2011,Torres2014} and references therein). However, none of them fully describe the spectral shape and the morphology observed with high precision: detailed studies
of the spectral shape of the VHE emission should allow us to finally disentangle the strength and structure of the magnetic field, together with the particle distribution function in the nebula.
The characterization of the spectral energy distribution (SED) peak is of special importance, since the SSC scattering becomes relevant only for the highly energetic ($\sim$70 percent of the Crab rotational power), particle dominated nebulae at low ages (of less than a few kyr), which are located in a FIR background with relatively low energy density \citep{Torres2013}.
It is particularly interesting the still poorly explored energy interval above few tens of TeV. The so-called \cutoff{} region is related to the maximum energy of the parent particles as well as to the energy losses, encompassesing significant information on particle acceleration and evolution. However, in the specific case of the Crab nebula, the spectral steepening at VHEs is a radiative or a propagation feature. Given the high magnetic field of $\sim$ 100 $\mu$G \citep{KC84,2010Meyer,Martin2012}, in fact, the most energetic parent particles dissipate their energies via synchrotron radiation. On one hand, the spectrum steepens due to the transition of the IC mechanism from the Thomson to the Klein-Nishina (KN) regime that occurs at different energies for the three dominant target photon fields: the synchrotron-self-Compton, the far infrared and the CMB \citep{1996Atoyan}. The overall spectral shape is, therefore, given by the overlap of the three contributions. On the other hand, also the energy-dependence of the particle diffusion mechanism can generate possible spectral breaks \citep{Lefa2012}. \\
Empirically, several analytical models were proposed to describe the VHE emission. It is already clear that there is no simple mathematical function that can properly describe the entire IC component, from $\sim$1\,GeV to tens of TeV, but there are good approximations in reduced energy intervals. At energies larger than a few hundreds of GeV, above the IC peak, the emission was fitted by a power law up to 80\,TeV (PL: \citealt{Aharonian2004}), a power law with an exponential \cutoff{} at 14.3\,TeV (PLEC-HESS, \citealt{Aharonian2006}), and a log-parabola (LP-MAGIC: \citealt{Aleksic2015}; LP-HESS: \citealt{Holler2015}, see also \citealt{2015Meagher}) reaching up to 100\, TeV \citep{Hawc2019}. Although it is, currently, commonly accepted that the Crab nebula spectrum exhibits a curvature above 10\,TeV, the exact position and shape of this spectral break is still under debate. In this work we test CTA capabilities to answer this question. 

The size of the Crab nebula results from the interplay between radiation losses and particle transport mechanisms and depends on the energy of the underlying particle population. The morphology of the synchrotron component has been studied in detail from radio to gamma rays (see for a review \citealt{Hester2008}), showing a clear dependence on the energy range.  
At X-ray energies the Crab exhibits a complex structure consisting of a torus and two narrow jets emerging from the direction perpendicular to the torus plane.  
The size of the Crab nebula is energy dependent as a consequence of the energy losses due to synchrotron burn-off, which limits the particle energy as a function of the distance from the shock, and therefore the size of the nebula.

The morphology of the IC nebula is, instead, poorly known due to the limited angular resolution of the gamma-ray telescopes which is of the order of 0.5\degr{} for the \fermi{} detector at few GeV and of few arcminutes for the current generation of IACTs. We expect its extension to be related with that of the synchrotron nebula, since it depends on the same population of the synchrotron electrons convolved with the spatial distribution of the photon field. 
At energies above 1\,GeV an extension of  0.03$^{\circ}$ \citep{Ackermann2018,Yeung2019} was derived using data from \fermi{}.
At higher energies the nebular shrinking due to synchrotron burn-off \citep{1996Atoyan} was finally established by the H.E.S.S. collaboration thanks to the use of advanced analysis techniques that improved the angular resolution down to 0.05\degr\ above 700\,GeV. The sigma of the measured Gaussian morphology is 52.2\arcs{}$\pm2.9$\arcs{}$_{stat}\pm7.8$\arcs{}$_{sys}$ \citep{2017HESSExtension}. Such extension is in good agreement with the theoretical expectations probing for the first time electron energies in the 1-10\,TeV range. This electron population is, in fact, responsible for the 0.1\,keV synchrotron emission that is inaccessible because of its absorption with the interstellar medium. 

The Crab pulsar is one of the few pulsars detected across the entire electromagnetic spectrum, from radio frequencies to $\gamma$ rays. At high energies, above 100\,MeV, the significant improvement in sensitivity and the unprecedented statistics afforded by \fermi{}, with respect to the previous generation of instruments, established precise measurements of the Crab pulsar spectrum. The phase-averaged spectrum is well represented by a 1.97 power law function with an energy cutoff at $(5.8 \pm 0.5 \pm 1.2)$\,GeV \citep{2010Abdo}, in agreement with the theoretical prediction of a synchro-curvature emission component \citep{2015Vigano} with a maximum energy limited by either magnetic and gamma pair absorption or radiation losses. In particular, the \fermi{} result, together with the MAGIC detection of the Crab pulsar at $\sim$25\,GeV \citep{2008Aliu}, suggested an emission originating at high-altitudes inside the magnetosphere, i.e. the outer gap models \citep{Cheng:1986}. 
Few years later, in 2011, the unexpected discovery of a new spectral component emerging above the cutoff at $\sim$6\,GeV, and extending up to hundreds of GeV opened a new window in the pulsar physics. Discovered by VERITAS \citep{2011ScienceVeritas}, and later confirmed by MAGIC \citep{2011Aleksic,2012Aleksic}, this results implies the existence of relativistic particles close to or beyond the light cylinder where absorption is negligible. It has been proposed that a new emission mechanism must be invoked to account for this new component, namely IC of lower energy synchrotron pulsed photons. The location of the emission, if magnetospheric \citep{2011Aleksic,Lyutikov:2012}, in the pulsar wind region \citep{2012NatureAharonian}, or in the current sheets \citep{Contopoulos:1999,Cerutti:2016} extending beyond the light cylinder \citep{2015Mochol} still remains an open question. Normal synchro-curvature emission (as it is the case for other pulsars, see \citealt{Torres2018}) could also play a role. The full (infrared to gamma-rays, with a continuous coverage) pulsed spectrum of Crab could be described with synchro-curvature radiation in a model using seven parameters (even less than needed to generate the Crab spectrum via a cyclotron self-Compton model \citealt{Lyutikov2013}). And it is not even discarded that to represent the TeV component with a small population of particles emitting towards us, but requiring a larger accelerating field (see figure 14 of \citealt{Torres2019} and the associated discussion). The presence of two different mechanisms for the production of the high-energy and the VHE components leads to the production of spectral features whose detection could shed light on the understanding of the mechanisms at work.  

The Cherenkov Telescope Array (CTA, \citealt{2013CTA}), currently in the pre-construction phase, is the next generation ground-based observatory for very-high-energy gamma-ray astronomy (up to more than 300\,TeV).
CTA will be located on two sites, a northern location in La Palma (Spain), and a southern one in Paranal (Chile). 
It will observe with an array of Small, Medium and Large Size Telescopes (SSTs, MSTs and LSTs),
improving the sensitivity of the existing VHE instruments by a factor of five to ten depending on the energy range.
The angular resolution of both arrays will be equal to or better than 0.1\degr{} at 0.1\,TeV and 0.05\degr{} at energies above 1\,TeV, and the energy resolution will be equal to or better than $30\%$ at 50\,GeV and $10\%$ at all energies above 1\,TeV (\citealt{2015Hassan}).

In this paper we aim at understanding the potential of CTA to study different characteristics of the Crab pulsar and nebula \citep{2013deOna,2016Burtovoi}, which have direct implications on the physics mechanisms behind the multi-wavelength radiation. In particular, ($1$) What is the maximum energy that CTA can achieve with deep observations of the nebula? or in other words, what is the largest-energy particles CTA can probe ($2$) How well can CTA constrain the nebula spectral shape, which depends on the particles acceleration and emission mechanisms at play? ($3$) Can CTA resolve the morphology of the Crab nebula? that is, how well can we study the transport and evolution of particles in a magnetised medium ($4$) What can we expect from deep observations of the pulsar with CTA? can we disentangle between different spectral components in the GeV/TeV regimes by means of CTA observations?  
In this work we only consider the Crab nebula in a steady state (out of a state similar to the several gamma-ray flares that have been observed \citealt{2011Abdo}).

It is important to note that the simulations showed in this paper are based in a conservative estimation of the CTA response functions, which are aimed to describe the general behavior of the instrument and based on standard Hillas reconstruction \cite{1998Hillas}.
To account for this, we also tested a more realistic and improved response modifying artificially some of the simulated telescope parameters. Thus, \S $2$ describes the Instrument Response Functions (IRF) of CTA, the modifications of those reflecting the foreseen improvement by using more sophisticated techniques, and the tools (\S $2.1$) we used in this study.
\S $3$ describes the hypotheses we assumed for this work and the results of the simulations regarding the Crab nebula spectral shape and morphology.
\S $4$ presents the results of the simulations concerning the pulsar and \S $5$ provides a few concluding remarks.
\section{Simulations and analysis}
We simulated the energy spectrum and the morphology of both the Crab nebula and pulsar. The simulations of both the spectral and the morphological behaviour require an a priori assumption on the model, which must be convoluted with the IRFs.

The IRF comprises the mathematical description that relates the observable of the events measured by the instrument (measured energy $E'$, measured incident direction $p'$ and arrival time $t'$), with the physical quantities of the incident photon (true energy $E$, true incident direction $p$ and true arrival time $t$).
It is factorized into the effective area $\rm{A}_{eff}(p,E,t)$, the point spread function $\rm{PSF}(p'|p,E,t)$, and the energy dispersion $\rm{E}_{disp}(E'\ |p,E,t)$. 
The factorization of the instrument response is given by the following expression:
\begin{dmath}
\rm{R}(p', E', t' \ | p, E, t)=\rm{A}_{eff}(p, E, t) \times \rm{PSF}(p' \ | p, E, t) \times \rm{E}_{disp}(E' \ | p, E, t)
\end{dmath}
In this work we used version $\rm{prod}3\rm{b}$ of the publicly available IRFs \footnote{https://www.cta-observatory.org/science/cta-performance/}. A detailed description on how these IRFs are obtained can be found in \citet{2015Hassan}. These IRFs have been optimized for the detection of isolated point-like sources against residual cosmic-ray background for exposure times of 0.5, 5 and 50\,hr and at two different zenith angles of 20$^{\circ}$ and 40$^{\circ}$.  We considered the IRFs for the Northern site at 20$^{\circ}$ zenith angle where the Crab culminates almost at zenith, thus being observable at small zenith angles which guarantee the lowest energy threshold possible. 
It should also be emphasized that the available IRFs do not account yet for the foreseen improvement that will be achieved by the use of advance analysis techniques (see for instance \citealt{2018arXiv181000592M} or \citealt{2019APh...105...44S}).
We expect that more advanced IRFs will improve significantly key parameters such the angular and energy resolution. To take this into account, we also evaluated the expected results assuming some degree of improvement beyond the current IRFs, as described in \S $3.2$.

\subsection{Analysis science tools}
\label{sec:ctools}
In this work we used both prototypes for the CTA Science Tools (STs): {\it gammapy} \citep{2017ICRCgammapy}  version $0.7$ and {\it ctools} \citep{2016ctools} version 1.5.0.

We implemented a simulation scheme within {\it Gammapy} that works as a generator of simulated $3\rm{D}$ sky cubes, which we have also made public\footnote[2]{https://github.com/emestregui/The-CTA-Crab-Nebula-and-Pulsar/tree/master/3D_cubes_simulator_archive}, with two spatial dimensions (sky direction) plus an energy axis discretised in bins.
These cubes contain the predicted counts for the simulated Crab nebula and pulsar, supposing that both sources are in the same position but with different morphology and spectra.
The set of simulated cubes comprise the predicted counts from the two sources (without background), an observation simulation with an ON and OFF cube, a cube of excess (ON minus OFF), and secondary products (also in the form of $3\rm{D}$ sky cubes) such as an exposure cube (effective area times observation time) or a cube with the total flux from the sources.
The ON cube comprises the predicted counts of both sources and background counts (being the three components smeared by Poisson noise).
The OFF cube contains only predicted background counts smeared by Poisson noise.
This is computed including the cosmic-ray background provided by the CTA IRFs for each particular offset (distance to the center of the field of view) and energy bin. 
The simulations take into account the CTA IRFs (effective area, energy resolution and the energy dependent point spread function) and the offset.

The sky cubes allow us to perform $1\rm{D}$ spectral analysis projecting the cubes along the energy axis, and $2\rm{D}$ morphological analysis using the {\it Sherpa} software \citep{2001Sherpa,SciPyProceedings} for each energy bin of the cube. 
{\it Gammapy} provides a $1\rm{D}$ {\it Sherpa} spectral fitting with maximum likelihood functions (Cash or Wstat, see \citealt{1979Cash}) statistics and Nelder-Mead Simplex optimization method \citep{NEADMED2,NEDMEAD1} based on a forward-folding technique (\citealt{2001Piron}). 
The input of the sky cube generator is an analytical spectral model (Table $1$) and a morphological model for each source together with the desired configuration of the cubes (spatial and energy axis binning, energy limits, pointing coordinates and cube size, coordinate system and projection, observation time and offset). 
We produced the cubes in the whole energy range of CTA covered by the IRFs, and observation times from seconds up to $300$ hours. 
The best-fit spectral parameters are derived following the procedure explained in 
{\url { http://docs.gammapy.org/dev/spectrum/fitting.html\#spectral-fitting}}.
%

The {\it ctools}\footnote[3]{http://cta.irap.omp.eu/ctools/} software package (\citealt{2016ctools}) has been developed to perform analysis of gamma-ray instruments. 
{\it ctools} is based on {\it GammaLib}. 
The spectral studies of the Crab nebula we present (\S $3.1$) have been checked using the two frameworks, {\it ctools} and {\it Gammapy}, independently.
To make this comparison we performed observation simulations of the Crab nebula and pulsar with {\it ctools} in addition to the simulated $3\rm{D}$ sky cubes. 
{\it ctools} performs maximum likelihood fitting of a model to unbinned or binned data, or joint maximum likelihood analysis in the case of data coming from different observations or instruments. 
We used maximum likelihood fitting in unbinned mode.
The results obtained with the two analysis chains were compatible within $1\sigma$ error.  

We performed simulations of the Crab pulsar (point-like source with the spectrum model detailed in \S $4$) for different observation times and fluxes in the energy range between $20$ GeV and $180$ TeV (with the mentioned CTA IRFs).
We used templates for different phase curve models, which are also explained in \S $4$. 
We assigned the events phases in the region of interest (defined as a circle of 0.1\degr{} of radius) according to the timing model from \citealt{2014Zampieri} at epoch 55178 MJD.
The {\it ctools} tool {\it ctprob} computes the probability for each event to either belong to the source or to the background component. 
This tool was used to reject the events with probability of belonging to the source smaller than $95$\% from the simulations. 
Once the time properties were applied to the events, the signal was evaluated fitting the individual peaks to normal distributions (applying maximum likelihood estimation) and the periodicity was analyzed with the H-test statistic (\citealt{1989DeJagger}).

\subsection{Systematic errors}
\label{sec:systematics}
The systematic errors are the dominant source of uncertainty when studying in detail a bright source such as the Crab. For this reason, we need to take them into account in our simulations to guarantee a correct assessment of the CTA capabilities.  

The reconstruction of events, the Monte Carlo determination of the effective area, and the uncertainty in atmospheric conditions and background are some of the many sources of systematic errors in air Cherenkov telescope measurements. 
The systematic error in the energy scale ($\rm{E}_{0}$), for example, is requested to not exceed $4\%$ \citep{CTA_sys}, which constitutes an improvement of at least a factor of $2$ over the current measurements \citep{Aharonian2004,Aharonian2006,Aleksic2015,Holler2015}.
We have taken into account that the systematic error of the flux (at a given apparent energy) must be less than a $10\%$ between $50$ GeV and $300$ TeV \citep{CTA_sys}. 
To include these errors on the spectral models considered in the Crab nebula studies (see \S $3.1$.), we first simulate the spectrum taking into account only statistical errors by performing 1D analysis fitting over the simulated data, obtained from the convolution of the analytical models (Table \ref{tab:modelssumary}) with the CTA IRFs.
The expected flux for each of the energy bins was then computed and smeared by adding quadratically the systematic errors to the statistical ones. The systematic error on the energy scale is already taken into account prior to the smearing of the expected flux in each energy bin. To include it, we allowed the energy scale to fluctuate according to a 4\% of systematic error, as discussed above. We refit the new spectral points, in which both the statistical error and the systematic ones are added, and we use these results as input for the discrimination tests (see \S $3.1$).
Finally, we have considered a factor $2$ improvement over the fiducial requirements for the systematic errors.

\section{The Crab pulsar wind nebula}

\subsection{Spectral shape}

\begin{table*}
\begin{threeparttable}
\begin{tabular}{cccccc}
\hline
\hline
& $\rm{N}_{0}$ & $\rm{E}_{ref}$ & $\alpha$ & $\beta$ & $\rm{E}_{\cutoff}$ \\
& $[\rm{cm}^{-2}\rm{s}^{-1}\rm{TeV}^{-1}]$ & [TeV] & & & [TeV] \\
\hline
PL & $2.83 \times 10^{-11}$ & 1 & 2.62 & - & - \\
\hline
LP-MAGIC & $3.23 \times 10^{-11}$ & 1 & 2.47 & 0.104\tnote{1} & - \\
\hline
LP-HESS & $17.9 \times 10^{-11}$ & 0.521 & 2.1 & 0.24 & - \\
\hline
PLEC-HESS &  $3.76 \times 10^{-11}$ & 1.0 & 2.39 & - & 14.3 \\
\hline
LP-HAWC & $2.35 \times 10^{-13}$ & 7.0 & 2.79 & 0.1 & - \\
\hline
\hline
\end{tabular}
\footnotesize
\caption{Spectral models for the IC component of the Crab nebula used in literature. The corresponding references, in order of appearance, are \protect\cite{Aharonian2004,Aleksic2015,Holler2015,Aharonian2006,Hawc2019}.}
\label{tab:modelssumary}
\begin{tablenotes}
\small
\item \tnote{1}  In \protect\cite{Aleksic2015} the log-parabola formula is written with decimal logarithm, in this work we used the natural logarithm instead.
\end{tablenotes}
\end{threeparttable}
\end{table*}

First of all, we estimated the minimum observation time that is needed to detect the Crab nebula at 5$\sigma$ level at different energy ranges. In particular, the highest energy bin (E > 50\,TeV) considered is the energy interval that will allow the definitive discrimination between the different above-mentioned spectral models. We simulated the Crab nebula spectrum from 20\,GeV up to 300\,TeV under the different spectral assumptions that are listed in Table \ref{tab:modelssumary}. The simulations are performed with {\it Gammapy} over equally spaced moving intervals of observation time spanning between $0.001$\,hr and $300$\,hr, with steps of increasing width within the $10^{-4}-0.1$\,hr. We computed the significance of each observation simulation using equation $17$ of \citet{1983LiMa}. We performed $5000$ realizations for each spectral model and observation time considered. Table \ref{tab:CTANorth} shows the minimum observation time needed to achieve a mean significance of $5\sigma$. In addition, it also shows that the Crab can be detected at 5$\sigma$ in less than 25 seconds for either model from Table \ref{tab:modelssumary} in the most sensitive among the computed bins [0.1-1]\,TeV. The highest energy bin containing a 5$\sigma$ signal in 300\,hr extends up to $E \approx 60$ TeV for the PLEC-HESS, and $E \approx 100$ TeV for the LP-MAGIC and the LP-HAWC assumptions. 
We estimated the CTA capabilities to disentangle between the different mathematical approximations listed in Table~\ref{tab:modelssumary}. 
We computed thus the expected distribution of the excess events in bins of energy for each of the considered spectral models and for $50$\,hr of observations. To evaluate the statistical uncertainties on the excess, each of the spectral models was simulated $5 \times 10^{3}$ times, and from these distributions the mean and deviance were obtained. We compared the distributions in pairs using a chi-square test. The probability of two distributions resulting from two different hypotheses to be compatible is rejected in all cases (with a probability of $95\%$ CL) when comparing the whole energy range of the simulations. This is mainly due to the differences of the hypotheses in the lower energy range. Even when considering only the \cutoff{} region, above 8\,TeV it is possible to discriminate between LP-MAGIC and PLEC-HESS at 95\% CL. Also, the LP-MAGIC, LP-HAWC and LP-HESS models were distinguishable among each other above 8\,TeV of energy. In addition, at energies above 50\,TeV the HESS-PLEC model was distinguished from MAGIC/HAWC log-parabolas and vice versa.

\begin{figure*}
\centering
\includegraphics[width=0.45\textwidth]{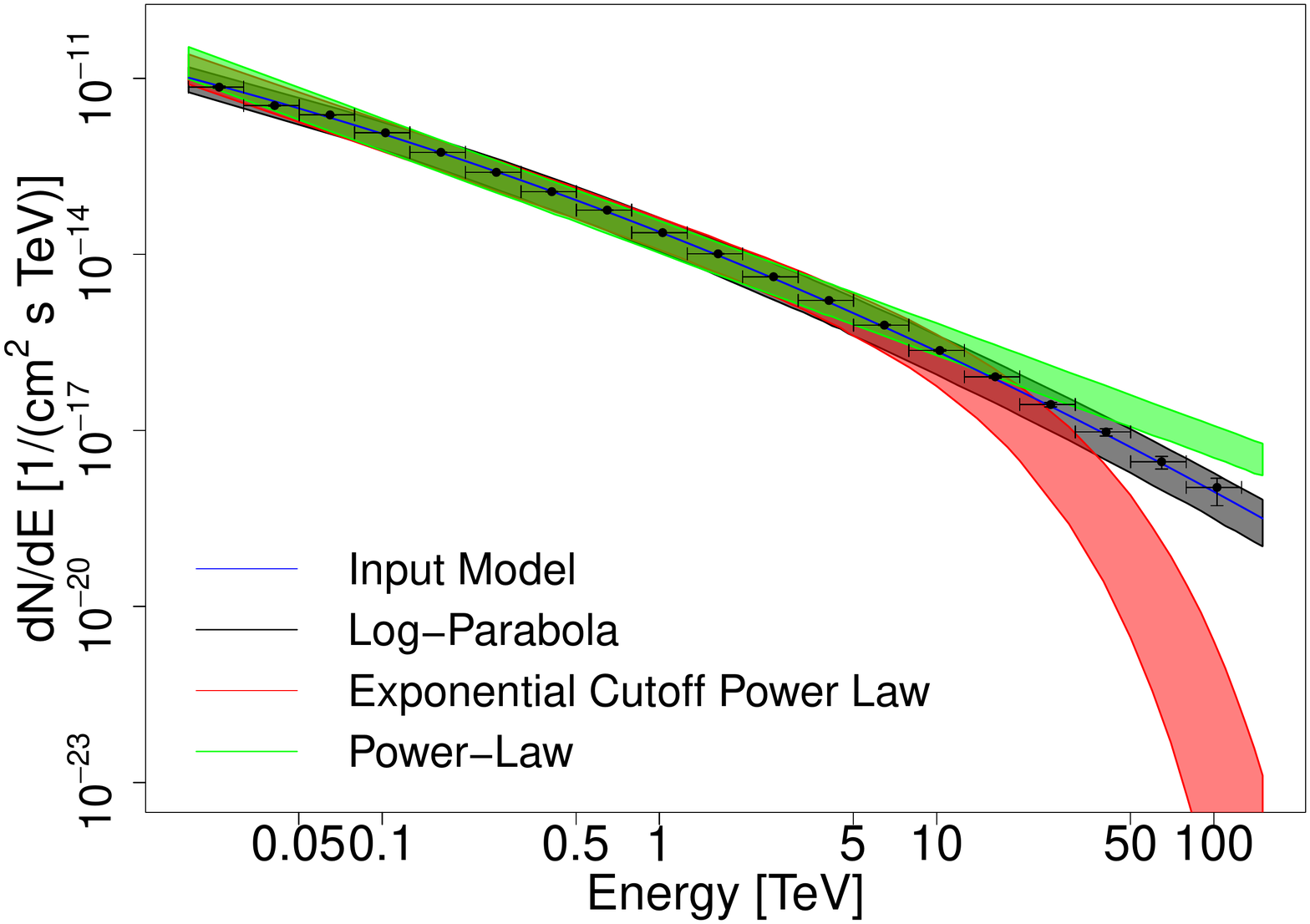}
\includegraphics[width=0.45\textwidth]{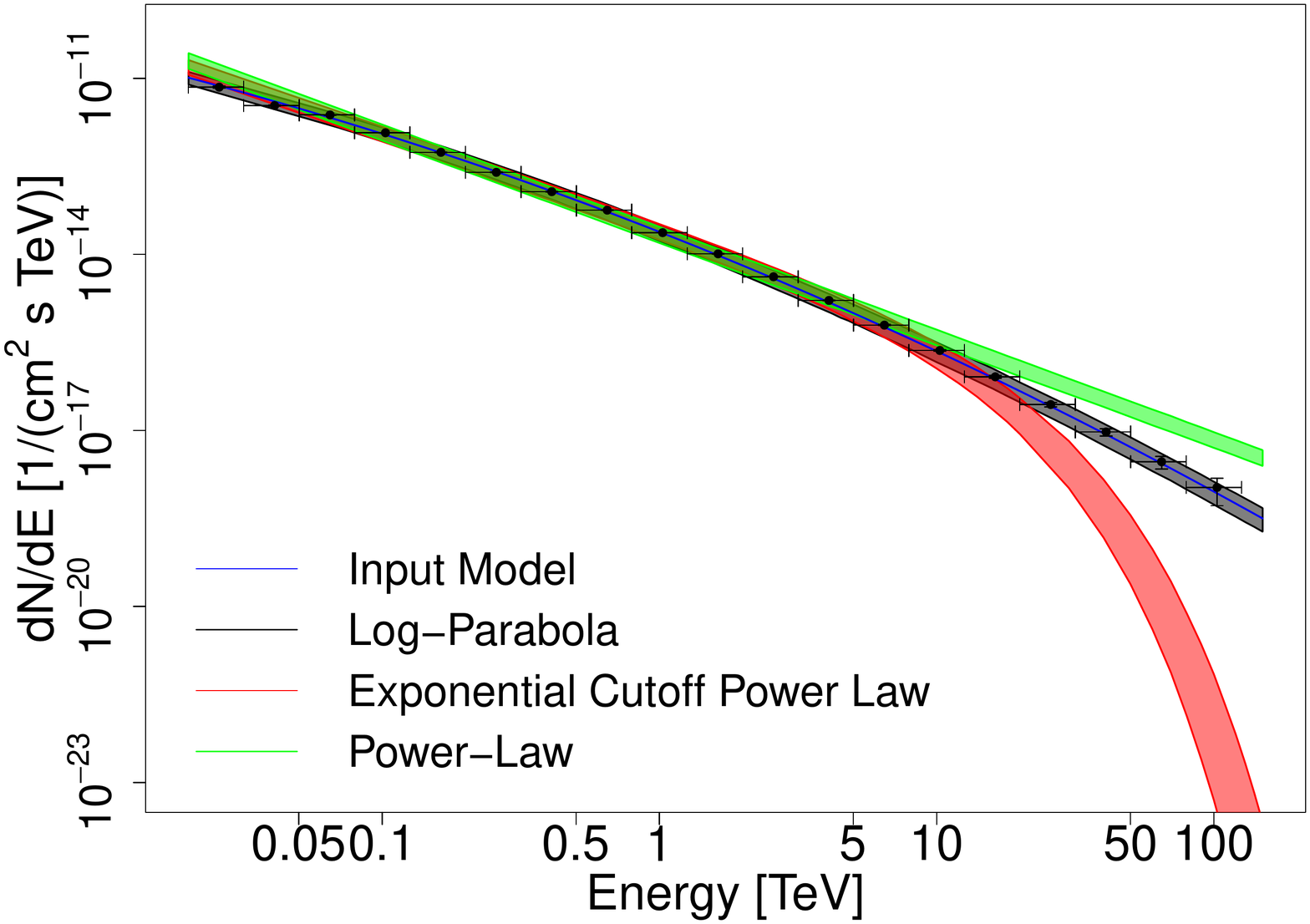}
\caption{Result of fitting all the spectral shapes considered to the simulations for the nebula of LP-MAGIC model (Table \ref{tab:modelssumary}), with $3\sigma$ error region noted. Results under the assumption of the systematic error requirements in \S  $2.2$ (left), and with the systematic uncertainty improved by a factor of $2$ (right) are shown.}
\label{fig:LP-MAGICspectrumfit}
\end{figure*}

We also fitted the data obtained simulating the LP-MAGIC spectrum for 50h of observation time to each of the spectral shapes listed in Table~\ref{tab:modelssumary} (log-parabola, power law and power law with an exponential cutoff). All the parameters of the models fitted to the data were free except for the energy of reference (fixed to 1\,TeV). The most energetic bin at which the source was detected was 50\,TeV - 80\,TeV  with a significance of $5\sigma$ (80\,TeV - 125\,TeV with a significance of $3\sigma$). We used the Akaike\textquotesingle s Information Criterion (AIC) \citep{Burnham2004} to evaluate the goodness of the fitted model. 
The AIC took a value of $33.7$ for the log-parabola model ($33.0$ if the systematic errors added to the simulated spectrum are improved by a factor $2$). For the power law and power law with an exponential cutoff fitted models, the value of the AICs were $84.4$ and $126.2$ respectively (or $190.2$ and $461.2$ respectively with the mentioned systematic improvement). Figure \ref{fig:LP-MAGICspectrumfit} shows the comparison between the models using the current systematic errors on the CTA requirements (left) and those ones improved by a factor 2 (right).

\begin{table}
\footnotesize
\caption{Minimum observation time (in hours) needed to obtain a $5\sigma$ detection of the Crab nebula in various energy ranges and under different assumptions for the spectral model. We considered the spectral models listed in Table~\ref{tab:modelssumary} }
\normalsize
\label{tab:CTANorth}
\begin{tabular}{p{0.2\linewidth}p{0.1\linewidth}p{0.1\linewidth}p{0.1\linewidth}p{0.1\linewidth}p{0.1\linewidth}}
 \hline
 \hline
 Energy [TeV] &  PL & LP-HAWC & LP-MAGIC & LP-HESS & PLEC-HESS \\
 \hline
 E $< 0.1$  & $0.027$ & $0.40$ & $0.37$ & $3.8$ & $0.059$\\
 $0.1 <$ E $< 1$ & $0.0053$ & $0.0037$ &   $0.0056$ & $0.007$ & $0.0046$ \\
 E $> 5$ &  $0.16$ & $0.13$ & $0.17$ & $0.22$ &  $0.15$\\
 E $>50$ & $10.1$ & $18.4$ & $30.3$ & $233$ & $298$ \\
 \hline
 \hline
\end{tabular}%
\end{table}

\begin{figure}
\centering
\includegraphics[width=0.5\textwidth]{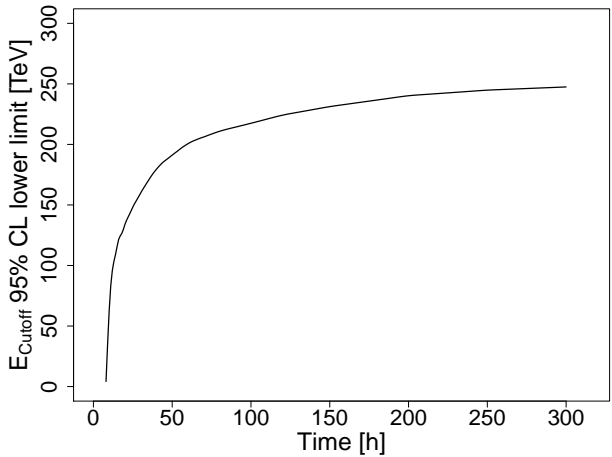}
\caption{Energy cutoff lower limit at $95$\% CL for different observation times.}
\label{fig:cutoff}
\end{figure}

Our second goal is to compute the lower limit on the energy \cutoff{} as a function of the observation time. We simulated a power law spectral model with the same $\alpha$ and N$_0$ value of the LP-MAGIC function for different observation times, spanning from 8\,hr to 300\,hr and $10^5$ realizations for each observation time, then we fitted the obtained results with a power law with an exponential \cutoff{} and computed the lower limit of the energy \cutoff{} at 95\% confidence level (CL). The results are illustrated in Figure \ref{fig:cutoff}.

\begin{figure*}
\centering
\includegraphics[width=0.45\textwidth]{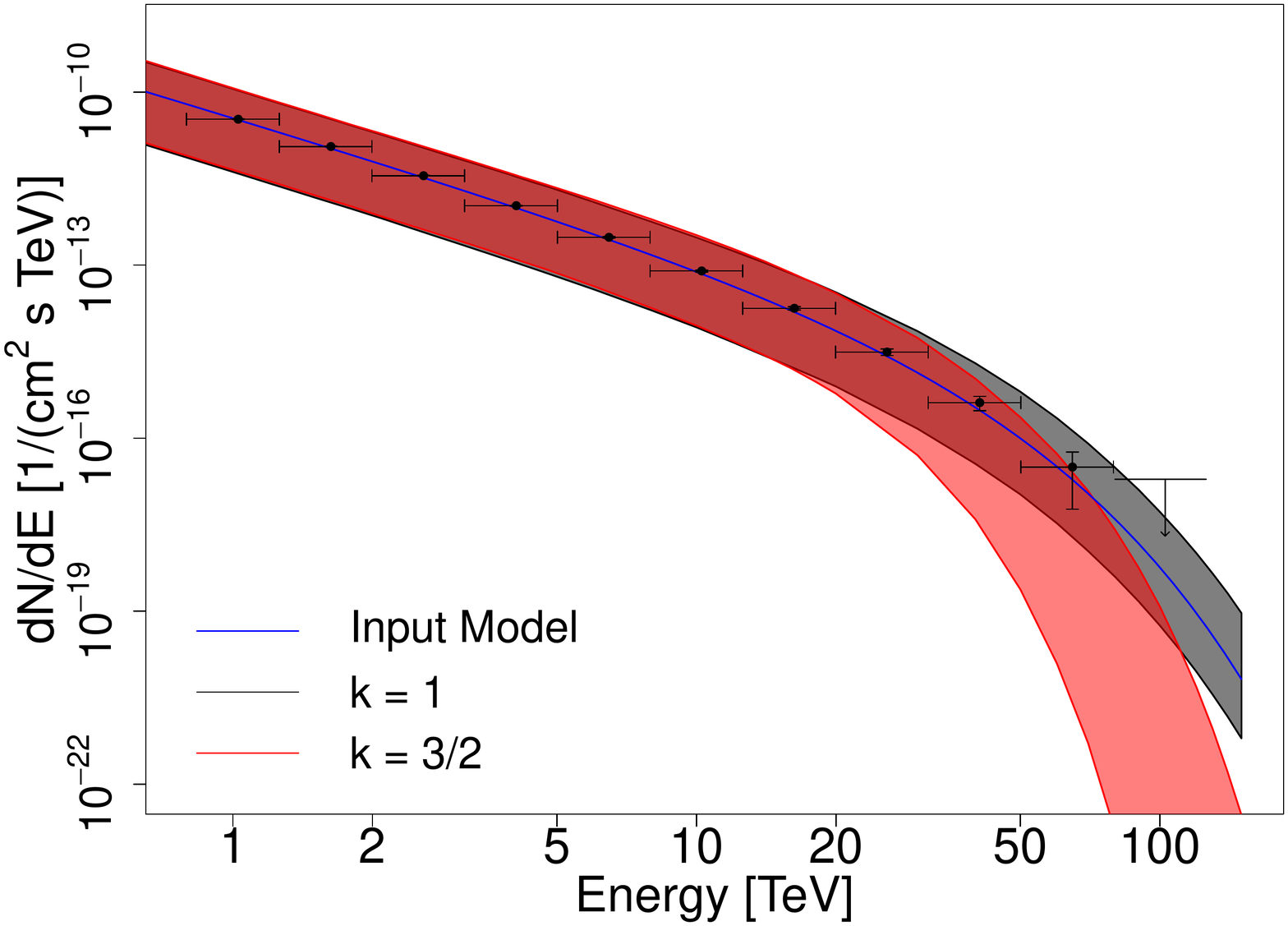}
\includegraphics[width=0.45\textwidth]{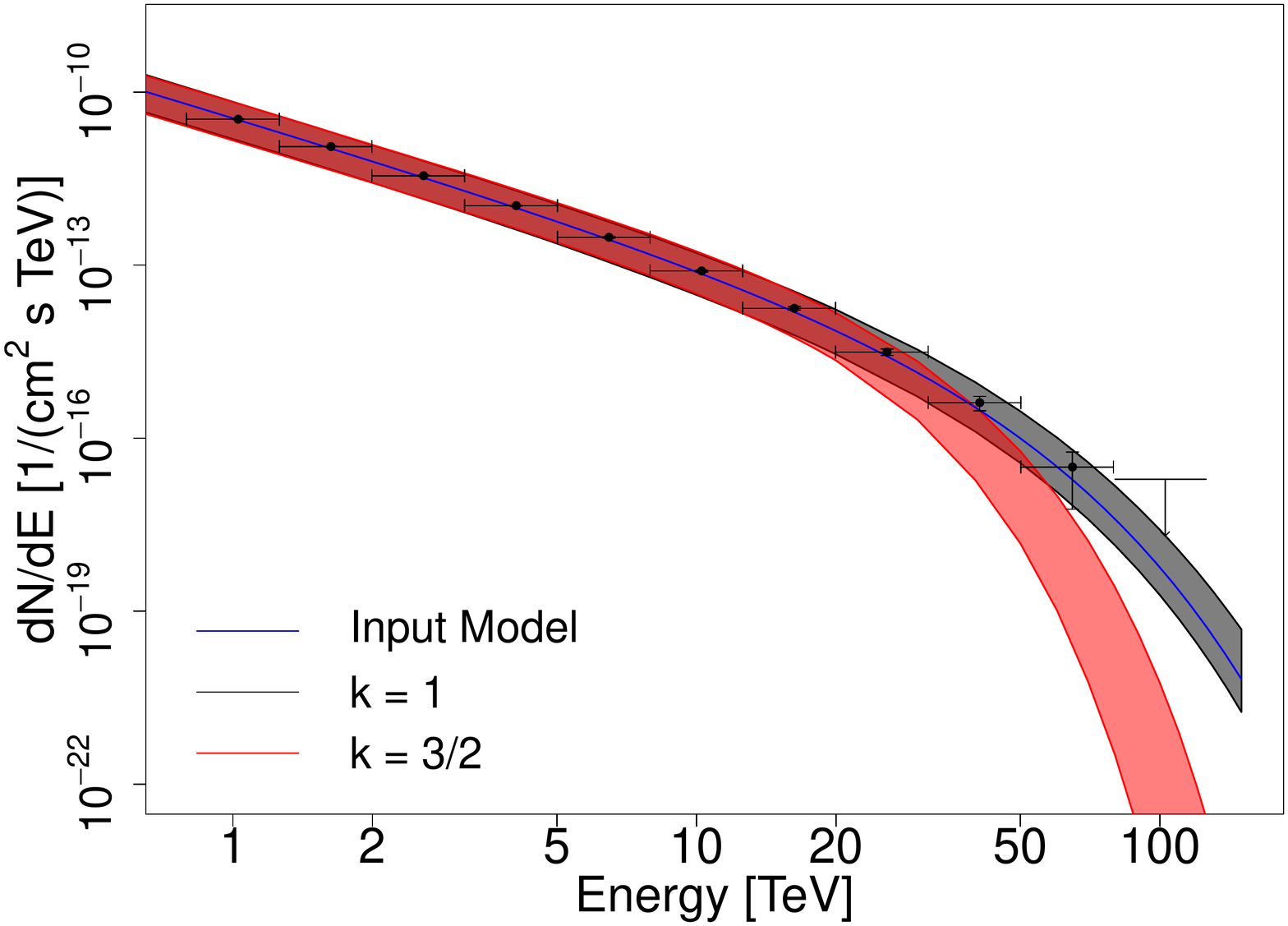}
\caption{CTA capabilities to disentangle between different shapes of an exponential spectral \cutoff, i.e. different values of $k$, as in Eq.~\ref{eq:pleck}. In particular, we show the two cases of $k$=1 and $k$=3/2. The shaded area represent the systematic and statistic errors, quadratically added. On the left, the errors are calculated using the systematic error requirements, whereas on the right, the systematic errors have been reduced by a factor $2$.}
\label{fig:pleck}
\end{figure*}

Finally, we test the CTA capabilities to disentangle between different \cutoff{} shapes, thus we considered the following modified power law with exponential cutoff function:
\begin{equation}\label{eq:pleck}
\dfrac{dN}{dE} = N_{0} \bigg( \dfrac{E}{E_{0}} \bigg)^{-\alpha} \mr{exp} \bigg[-\bigg( \dfrac{E}{E_{\rm{cutoff}}} \bigg)^{k} \bigg]
\end{equation}
Different values of $k$ can reflect either the energy-dependence of the particle propagation mechanisms with $k$=1/3, 3/11, 1/4, and 1/5 being the Bohm, Kraichnan, Kolmogorov, and hard-sphere turbulence spectral models (in Thomson regime), respectively \citep{Romoli2017} or the transition from the Thomson to the KN region of the IC mechanism.
In the specific case of IC scattering of synchrotron photons in the KN regime, which is the dominant regime in the Crab nebula, \citealt{Lefa2012} computed the following possibilities: $k$= 3/2, 5/3, 2, 3 (corresponding, in this regime, to the same electron cutoff shape; $k = \beta_{e}$, see Table 1 of the cited paper). We assumed as primary hypothesis the PLEC-HESS spectral model which is the $k=1$ specific case of Eq.~\ref{eq:pleck}. This can be considered as a rather optimistic case, since the latest results by MAGIC \citep{Aleksic2015}, H.E.S.S. \citep{Holler2015} and HAWC \citep{Hawc2019} indicate that the spectral break should occur at higher energies, where the statistics becomes more and more an issue. We simulated a power law with an exponential cutoff spectrum following the PLEC-HESS assumption for 50\,h of observation time and then we fitted the results with Eq. \ref{eq:pleck} and $k$=1, 3/2, 5/3, 2, 3. The AIC computed for each of these fitting functions with respect to case of $k$=1 (taking the values of 34.6, 52.7, 79.6, 196.31 for $k$=1, 3/2, 5/3 and 2, and a value $>> 200$ for $k$=3) is always positive and indicates that it is possible to disentangle between an exponential and a super-exponential cutoff shape with a minimum $k$ value of 3/2 (see Figure \ref{fig:pleck}). We also simulated the various Eq. \ref{eq:pleck} using the above $k$ values to be compared against each other. The results show that it will be possible to discriminate within each other, even when including systematic uncertainties. The smallest difference in the $\beta$ parameter that was distinguished at 95\% CL from the $\beta = 1$ model (for the observation time and systematic error improvement considered) was $\Delta \beta \approx 0.2$.

The comparison of the fitting results obtained with the two science tools is shown in the appendix. The results obtained with the two different science tools are consistent within the statistical uncertainties. 

\color{black}

\subsection{Morphology}

In the following, we study the capabilities of CTA to measure the nebular shrinking with increasing energies, hence to determine what is the maximum energy at which the Crab can be significantly resolved by CTA. In addition, we will assess the potentialities of CTA in detecting asymmetries in the Crab morphology. The Crab at ten TeV is expected to exhibit the jet-torus structure revealed by \emph{Chandra} \citep{2008Volpi} and although the arcsecond resolution of the X-ray detectors is certainly beyond the CTA performance, such a complex structure translates on a clear asymmetry in the two spatial dimensions. 

For all simulations performed in this section we accounted for 50 hr of observations and we assumed a spectral shape following the prescription of the LP-MAGIC model. 

\begin{figure}
\centering
\includegraphics[width=0.5\textwidth]{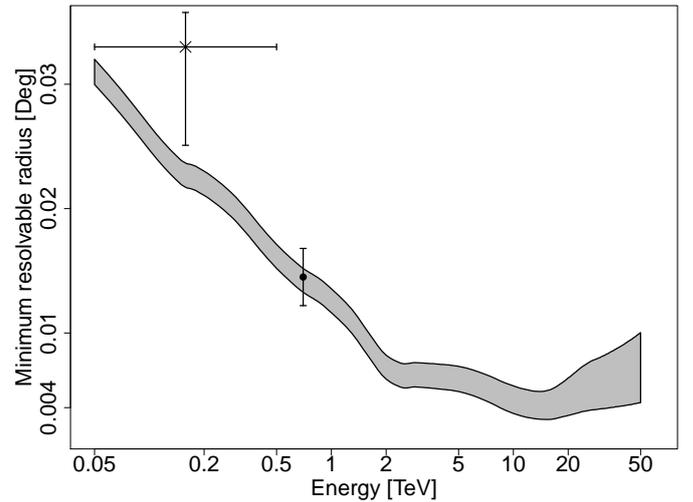}
\caption{Minimum size resolvable in the simulations versus the lower energy bound of the energy interval simulated (the upper bound is always fixed to 300\,TeV), using the CTA IRFs with the $1\sigma$ region noted. The measured size for the Crab nebula with H.E.S.S. for energies above 0.7\,TeV (black dot), and the one measured from 5\,GeV to 500\,GeV with \fermi{} (black star) are plotted with the $1\sigma$ error bars for comparison.} 
\label{fig:Minimumsizeresolved}
\end{figure}

To evaluate the capabilities of CTA to resolve the extension of the Crab nebula, we simulated in {\it Gammapy} 3D sky cubes a 2D projected spherical source with a LP-MAGIC spectral model. We varied the sphere radius from 0.004\degr{} (the smallest value of size per bin considered in the simulated sky cubes was of 0.002\degr{}/bin) to 0.03\degr{} in steps of 0.002\degr{} for 14 different integral energy bins of lower energy bounds equally spaced in a log-scale spanning from 50\,GeV to 50\,TeV and a fixed upper bound of 300\,TeV. We assume a Gaussian-shaped PSF and obtained the $68$\% containment radius from the CTA IRFs. After convolving the energy-binned sky maps with the corresponding PSF, we fit the resulting image with a projected 2D sphere convolved with the PSF using the sherpa tool. We called the minimum resolvable radius at a given energy bin to the smallest radius of the source for which the fitted size is significantly greater than zero (at $3\sigma$) and also compatible with the input radius at $95$\% CL. The minimum resolvable radius is shown in Fig. \ref{fig:Minimumsizeresolved}, which reflects the effect of the PSF and the spectral shape. The cross and dot mark respectively the extension of the Crab nebula as measured with \fermi{} and \hess{} The results show that CTA will be able to resolve the Crab nebula at a large energy range, from tens of GeV to TeV energies.

\begin{figure*}
\centering
\includegraphics[width=1.\textwidth]{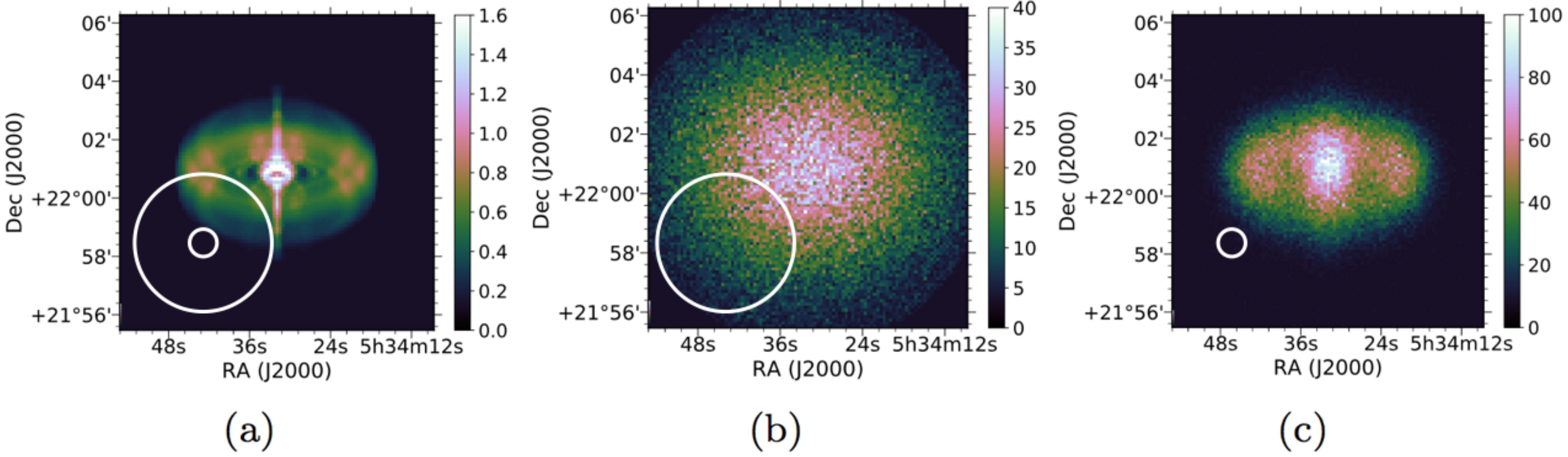}
\caption{a) Morphological template used for the jet-torus scenario. The white circles represent the PSF one sigma for the full system requirements (of approximately 0.04\degr{} for the energy bin spanning from 0.8\,TeV to 1.25\,TeV), and the same improved by a factor of $5$ (i.e. 0.008\degr{} referred to the same energy bin). It is a surface-brightness map in units of $\rm{erg}\ \rm{cm}^{-2}\rm{sr}^{-1}\rm{s}^{-1}\rm{Hz}^{-1}$. b) Observation simulation of 100 hours with the CTA full system requirements for the angular resolution plotted with the corresponding PSF (white circle), in counts (after background subtraction). c) Observation simulation as in b), but with the angular resolution improved by a factor of $5$ (exceeding by far the current CTA system requirements), again with the corresponding PSF.}
\label{fig:masym}
\end{figure*}

To prove the capabilities to detect asymmetries in the morphology of the Crab at different energies, we used as spatial template for our simulations the synthetic surface brightness map at 1\,TeV resulting from the 2-dimensional MHD simulations of the Crab performed in \citet{2008Volpi} (see Fig. \ref{fig:masym}a). To perform this simulation, we used {\it ctools} to import the 2D fits image provided by \citealt{2008Volpi} and the CTA North full system requirements for the angular resolution (which are in good agreement with the angular resolution provided by the version of the IRFs used for previous analysis). The resulting sky map is shown in Fig. \ref{fig:masym}b for 100\,h of observation time and an energy range spanning from 0.7\,TeV to 100\,TeV. To test the asymmetry of the resulting simulated events maps we performed 2D {\it Sherpa} morphological fittings. We compared the fitting results obtained by using a symmetric 2D Gaussian function and an asymmetric one. In the first case, a reconstructed size of (0.017(8) $\pm$ 0.0002)\degr{} is obtained, whereas the asymmetric function is favored (with a $\sqrt{\rm{TS}} \simeq 21$) resulting in a $\sigma_{\rm x}$ of (0.021(1) $\pm$ 0.0002)\degr{} and a $\sigma_{\rm y}$ of (0.013(8) $\pm$ 0.0003)\degr{} along the major and minor axis respectively.
 For illustrative purposes, we improve artificially the angular resolution by a factor 5, to show that in this case a jet-torus structure would be easily established (see Figure \ref{fig:masym}c, in fact a hint of the torus shape is already visible with a factor $4$ of improvement). Also, in this case, the error in the reconstructed major and minor axis ($\sigma_{\rm x}$ and $\sigma_{\rm y}$) for an assymetric Gaussian fit is improved by a factor $\sim 10$, retrieving an asymmetry of $\sigma_{x} / \sigma_{y} \sim 1.31 \pm 0.03$.

\section{The Crab pulsar}

The state-of-the-art measurements of the Crab pulsar at VHEs show a three-feature light curve with two peaks, the main pulse P1 and the interpulse P2, and an extra emission component between P1 and P2 dubbed bridge emission. P1, which is the main pulse at GHz frequencies, defines the phase 0, whereas P2 is shifted by $\sim 0.4$ with respect to P1. Both width and intensity of the pulses, as well as of the bridge emission, are energy-dependent \citep{kuiper2001}. In particular, in gamma rays, P2 becomes the dominant emission component at around 50\,GeV \citep{2014Bridge}, and both pulses are much narrower than in the \fermi{} energy band. In this work we assume as definition of the phase intervals the one in \citet{2012Aleksic} that is based on the VHE light curve fit results: $[-0.017-0.026]$, and $[0.377-0.422]$ for P1 and P2, respectively and the range in between for the bridge emission.   
The energy spectrum of the three components is well described by power law functions from 10\,GeV on with P1 ($\Gamma = 3.5 \pm 0.1$) being $0.5 \pm 0.1$ steeper than P2 \citep{2016Ansoldi}.  
MAGIC reconstructed spectral points of the harder and brighter interpulse up to $\sim$1\,TeV, with a lower limit on the energy of a possible cutoff at $\sim$700\,GeV \citep{2016Ansoldi}. The bridge emission, that is very prominent at the energies accessible by \fermi{}, is well represented by a 3.35 power law function above 50\,GeV and fades away at 150\,GeV \citep{2014Bridge}.

We simulated in {\it Gammapy} 3D sky cubes a point-like source at the position of the radio Crab pulsar (RA = 05:34:31.992 and DEC = +22:00:51.84, \citealt{2011A&A...533A..10L}) from 20\,GeV to 180\,TeV of energy and with a spectral index of 2.9 for a power law spectral model of $2 \times 10^{-11}\ \rm{cm}^{-2}\rm{s}^{-1}\rm{TeV}^{-1}$ of flux at 150\,GeV \citep{2016Ansoldi}. 
We included in our background model the continuous emission from the Crab nebula, that was simulated as an extended source with the LP-MAGIC spectral shape. The maximum energies at which the source was detected for 300\,hr of observation time were achieved in the energy bin spanning from 3\,TeV to 7\,TeV, at $3\sigma$.
Note that in the spectral studies performed for the Crab pulsar, the phase-average emission was simulated considering the spectrum of P2 \citep{2016Ansoldi}, which is harder than P1. Therefore, our approach to the observation simulations of the Crab pulsar can be considered as a somewhat optimistic case.

\begin{figure}
\centering
\includegraphics[width=0.5\textwidth]{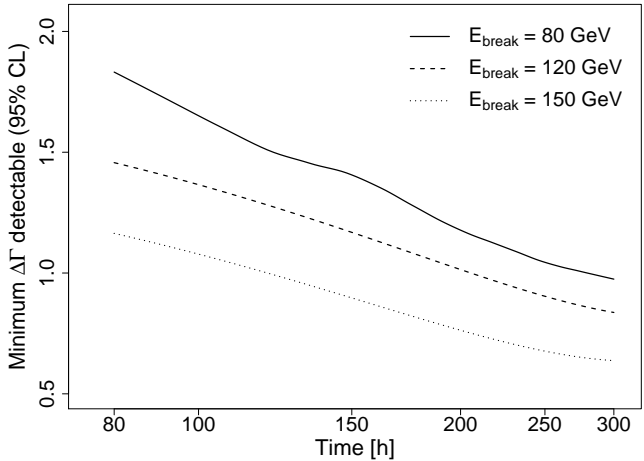}
\caption{Minimum index variation detected (at $95$\% CL) resulting from fitting the pulsar spectrum to a smoothed broken power law, fixing different energy breaks (in the legend) at different observation times.}
\label{fig:minimumindex}
\end{figure}

To explore the presence of spectral features, for example, a `kink' at $\sim$  100\,GeV resulting from two different spectral components we fitted the simulations of the pulsar to a smoothed broken power law fixing different values for the energy break, ranging from 80 to 150\,GeV. From the fitted models, we retrieve the precision on the measurement of the spectral indices below and above the break energy assumed. The minimum difference between the two indices for which the total spectrum is no longer compatible with the single power-law input function at $95$\% CL, is defined as minimum index variation detectable. We obtained that the minimum index variation that could be detected was of $\sim$ 0.6 at $95$\% CL in $300$ hours with the energy break located at 150\,GeV (see Fig. \ref{fig:minimumindex}). The minimum index variation detectable improved as the break was located at a higher energy, in part because of the improvement of the CTA sensitivity and the reduction of noise, but mostly because the spectrum fit below the break (down to 20\,GeV of energy) improves rapidly with the enlargement of the fitted energy range (and therefore with the increasing amount of energy bins to fit). The spectral fit above the break, however, is less sensitive to the energy break location, since the fitted energy range can be extended up to TeV energies. Note that the spectral index below the break energies considered can be better constrained combining CTA with measurements at energies below 20\,GeV performed with other instruments (as e.g. MAGIC and/or VERITAS). The minimum index variation detectable above the break energies with respect to the input model index (of 2.9) corresponds approximately to the one in Figure \ref{fig:minimumindex} improved by a factor $0.5-0.7$, depending on the break location and observation time.

\begin{figure*}
\centering
\subfigure[]{\includegraphics[width=0.45\textwidth]{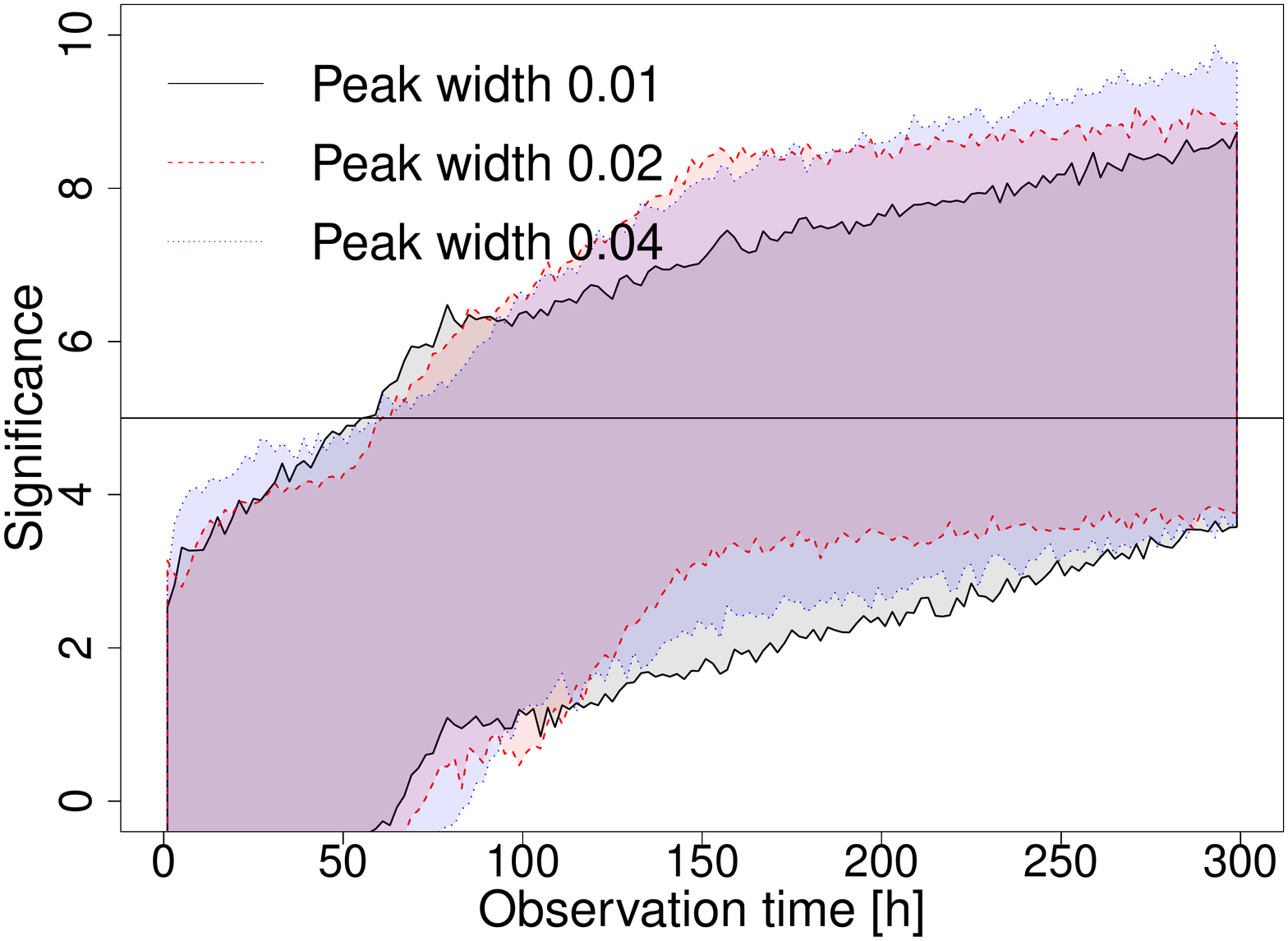}}
\subfigure[]{\includegraphics[width=0.45\textwidth]{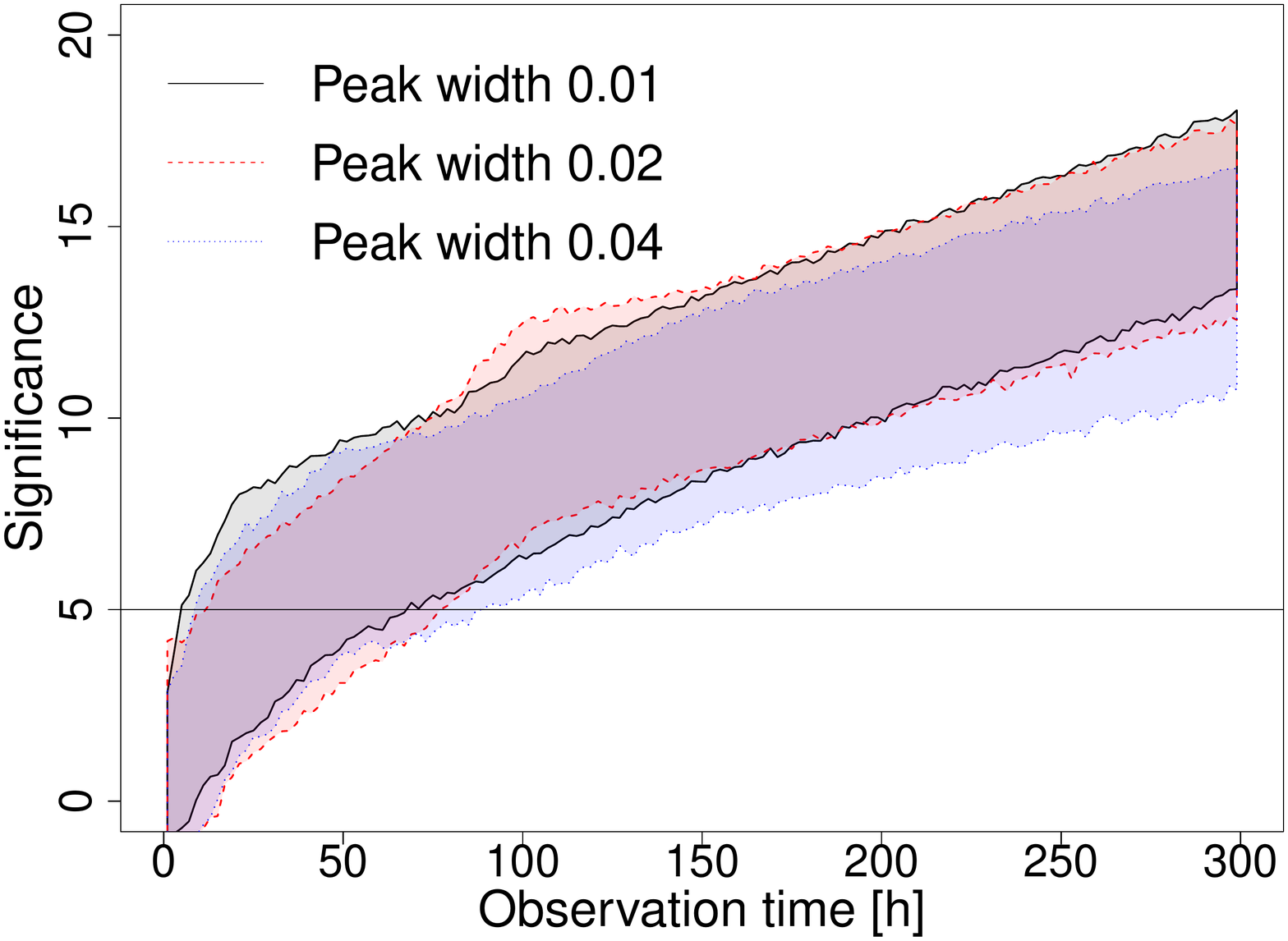}}
\subfigure[]{\includegraphics[width=0.45\textwidth]{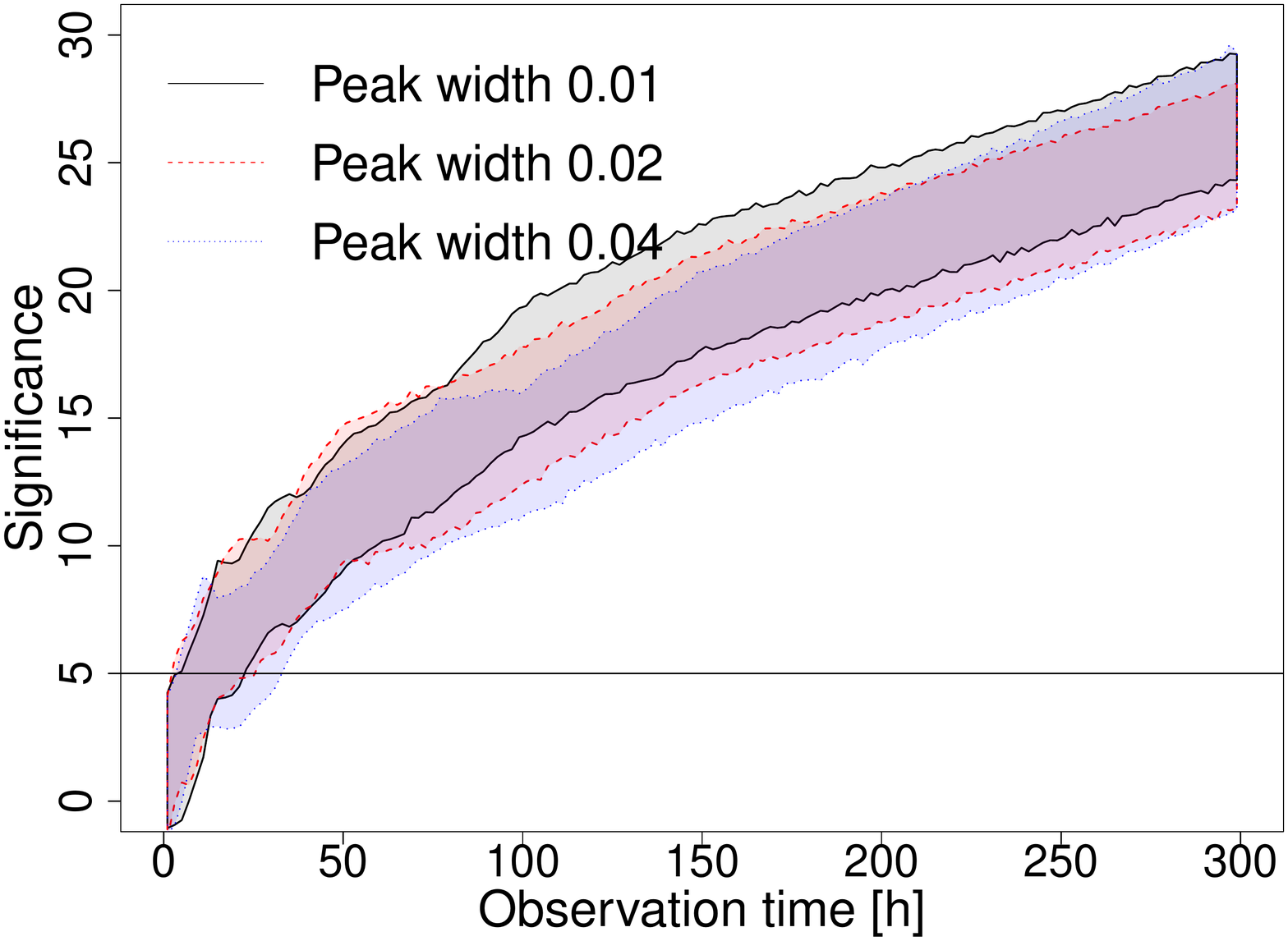}}
\subfigure[]{\includegraphics[width=0.45\textwidth]{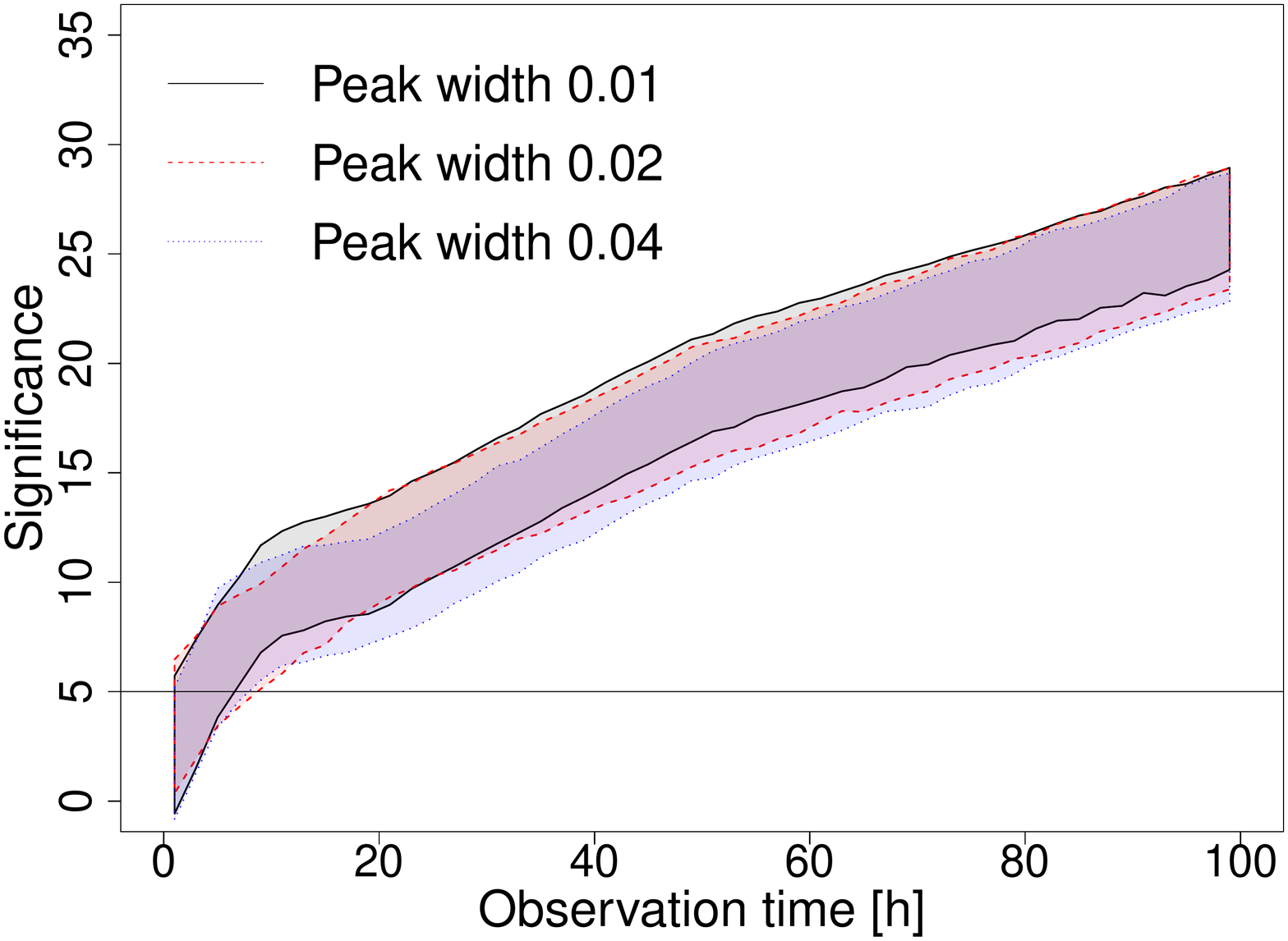}}
\caption{Evolution of the significance of the periodicity detection ($3\sigma$ region) with time for a source of flux (a) $0.2$, (b) $0.5$, (c) $1$, (d) $2$, times the integral flux of the Crab pulsar from 20\,GeV to 180\,TeV of energy and the spectrum considered}. The curves are smoothed by interpolation method and the $3\sigma$ region was computed by bootstrapping (\citealt{1979Efron}) of the H-statistic and simulations. 
\label{fig:periodicitysignificance}
\end{figure*}

\begin{figure}
\centering
\includegraphics[width=0.5\textwidth]{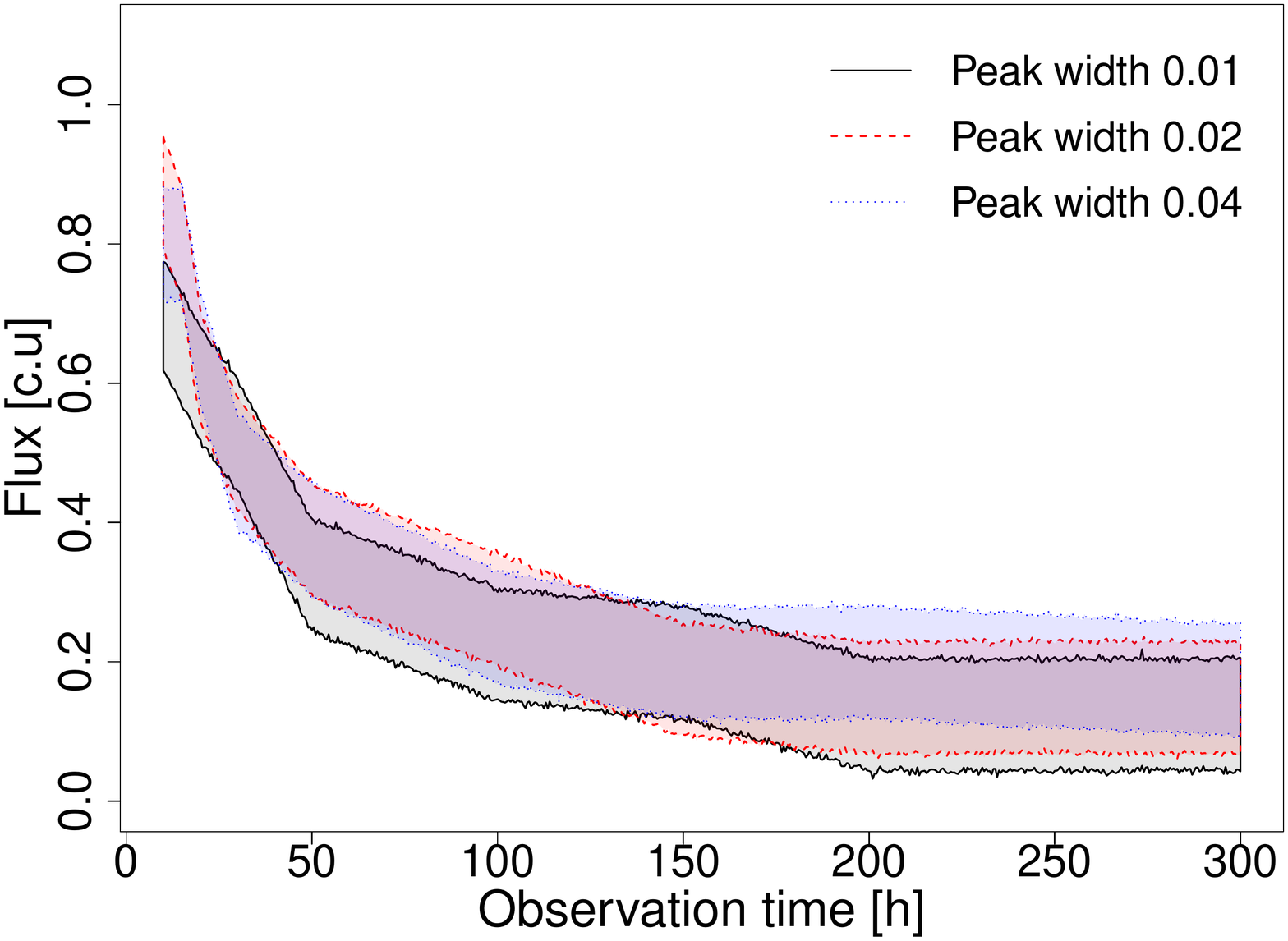}
\caption{Minimum source flux (in the same units of Figure \ref{fig:periodicitysignificance}) necessary to detect a periodicity at a $5\sigma$-level as a function of the observation time for Gaussian peak templates. The shadowed regions depict the $3\sigma$ error. }
\label{fig:Minimumfluxsignificance}
\end{figure}

Finally, we extended the simulations to study the capability of CTA to detect pulsed emission, considering different light curve shapes. For that, we used in {\it ctools} different templates with a single Gaussian peak (discretised in bins of 0.01 of width in phase) in the position of $\rm{P}2$ and gaussian $\sigma$ of $0.01$, $0.02$ and $0.04$ in phase, at different levels of flux. 
The periodicity was proved with the H-test statistic for the events with folded phases (\citealt{1989DeJagger}), and the probability associated with an H-test statistic for the best number of harmonics was taken from \cite{2010DeJagger} result and converted into a Gaussian sigma (with the area under the Gaussian probability density function integrated from $\sigma$ to infinity). 
The results are shown in Figure \ref{fig:periodicitysignificance}. 
We also tested the minimum flux necessary to detect the periodicity at $5\sigma$, fixing the observation time (Fig. \ref{fig:Minimumfluxsignificance}). Figures \ref{fig:periodicitysignificance} and \ref{fig:Minimumfluxsignificance} then represent the sensitivity of CTA to detect pulsars displaying different light curves and flux levels above 80\,GeV, and can be used for future pulsar population studies. 

\section{Discussion and conclusions}

We have studied the expected capabilities of the future CTA observatory regarding the observation of the Crab nebula and pulsar, testing the main existing tools for CTA simulations and data analysis, {\it Gammapy} and {\it ctools}. 
Note that this study is limited in some respects; the CTA IRFs used for this work are not optimized for the specific simulations described in this paper, and do not take into account the IRFs of the southern CTA telescope array, which is expected to optimize the spectral results at high energy. In addition, the analysis has not been optimized for angular resolution.
However, from this first approach and considering the limitations of the available simulations, we can derive the following statements: 
\begin{itemize}
\item (1) Our approach to the problem, the $3\rm{D}$ {\it Gammapy} sky cube generation allow us to place simulations of two objects in the same position with different morphology and spectral characteristics, plus the background. Therefore, this tool can be used for other complex systems or to study other important problems like source confusion in crowded regions of the sky. We also tested the two main analysis chains developed by the CTA Consortium, {\it ctools} and {\it gammapy}, and demonstrated they produce statistically compatible simulations and results of the spectral analysis.   
\item (2) We applied a statistically approach to estimate the CTA potential to answer all the open questions with respect the gamma-ray emission attributed to the Crab Nebula. We also included the, usually overlooked, systematic errors. Even considering the latest, which are the largest source of uncertainties in the current measurements, we showed that CTA will be able to discriminate between different hypotheses in a moderate observation time. Given the large energy range covered by the detector, CTA will disentangle between the different mathematical descriptions proposed (log-parabola, exponential cutoff power-law and simple power-law) within less than 50\,h with high confidence. We estimated not only the minimum cutoff that can be proved, but also the ability to parametrize the shape of this cutoff within a precision of $\Delta\beta = 0.2$, CTA will determine the domain in which particles are accelerated and cooled (see \citealt{Lefa2012}).  
\item (3) Concerning the morphology, we derived a general shape for the minimum size detectable that can be easily scaled when more realistic IRFs are used. Within this conservative approach, we already proved that CTA will provide invaluable input when comparing with theoretical predictions, for instance as a result of hydrodynamical or magneto-hydrodynamical simulations. We proved that CTA will be able to disentangle sub-parsecs structures ($\sim 0.02$\degr{} for a distance of $\sim 2.2$ kpc), which is crucial to understand the origin of the emission observed. In fact, our simulation shows that CTA will be able to image the nebula for the first time, allowing for instance the comparison with high energy resolution images in the X-ray regime. The simulations also show that CTA will access the effect of cooling processes of the electrons powering the nebula (see Fig. \ref{fig:Minimumsizeresolved}), and provide a continuous resolved image from a few tens of GeV to tens TeV. This will close the gap between the size measured by \fermi{} and the one by H.E.S.S. \citep{2017HESSExtension}, tracing the behavior of multiple-energy electron populations.
\item (4) With respect to the Crab pulsar, we studied the energy range in which the pulsed spectrum can be measured, and therefore the maximum energy to which the particles are accelerated. CTA will be able to detect pulsed photons up to 7\,TeV in a time of 300\,h. Even if this observation time (which is nevertheless conservatively estimated) looks high, it is important to mention that Crab is considered the standard candle in the gamma-ray domain, and therefore will be deeply observed from the beginning of the observatory working time. If the spectrum extends indeed well beyond 1 \,TeV, some of the models proposed to explain the extension of the pulsed emission, such as the ones based on synchrotron emission, will be heavily challenged. Likewise, the maximum energy detected will provide direct evidence of the region in which the radiation is located and therefore on the inner structure of the pulsar wind nebula. Also, the minimum spectral index variation that could be detected if an energy break occurs between 80\,GeV and 150\,GeV was established. Finally, we derived a general description of the power of CTA to detect different light curves shapes (from sharp peaks as in Crab, $0.01$ gaussian sigma width in phase, up to $4$ times wider peaks). 
Measuring with high-quality data the shape of the VHE pulse profile and comparing it with those obtained at lower energies can be used for constraining the location of the emission regions in different bands.
\end{itemize}

Certainly, our knowledge of the instrument response will increase in the upcoming months, during the construction and scientific verification phase of the observatory. The results summarised in this work are a conservative estimation of the potential of CTA to unveil the many unknowns behind the physics of the Crab Nebula and pulsar gamma-ray radiation. More realistic studies will emerge from the science verification phase describing the real power of CTA.

\section{Acknowledgements}

This research was supported by the grants AYA2017-92402-EXP, SGR2017-1383, PGC2018-095512-B-I00, SGR2017-1383 and iLink 2017-1238.
This research made use R Project for Statistical Computing (\citealt{Rmanual}). 
This research made use of the CTA instrument response functions provided by the CTA Consortium and Observatory, see http://www.cta-observatory.org/science/cta-performance/ (version prod3b-v1) for more details. 
This research made use of {\it ctools}, a community-developed analysis package for Imaging Air Cherenkov Telescope data. 
{\it ctools} is based on {\it GammaLib}, a community-developed toolbox for the high-level analysis of astronomical gamma-ray data. 
This research also made use of {\it Astropy}, a community-developed core Python package for Astronomy ({\it Astropy} Collaboration, 2018).

We are very thankful to D. Volpi, L. Del Zanna, E. Amato and N. Bucciantini for providing us with the synthetic surface brightness map of Crab at $1\rm{TeV}$ published in \cite{2008Volpi}. E. de O. W. acknowledges the Alexander
von Humboldt Foundation for financial support. 

\section{Appendix}

\subsection{Comparison of Gammapy and ctools results}

\begin{figure*}
\centering
\includegraphics[width=0.45\textwidth]{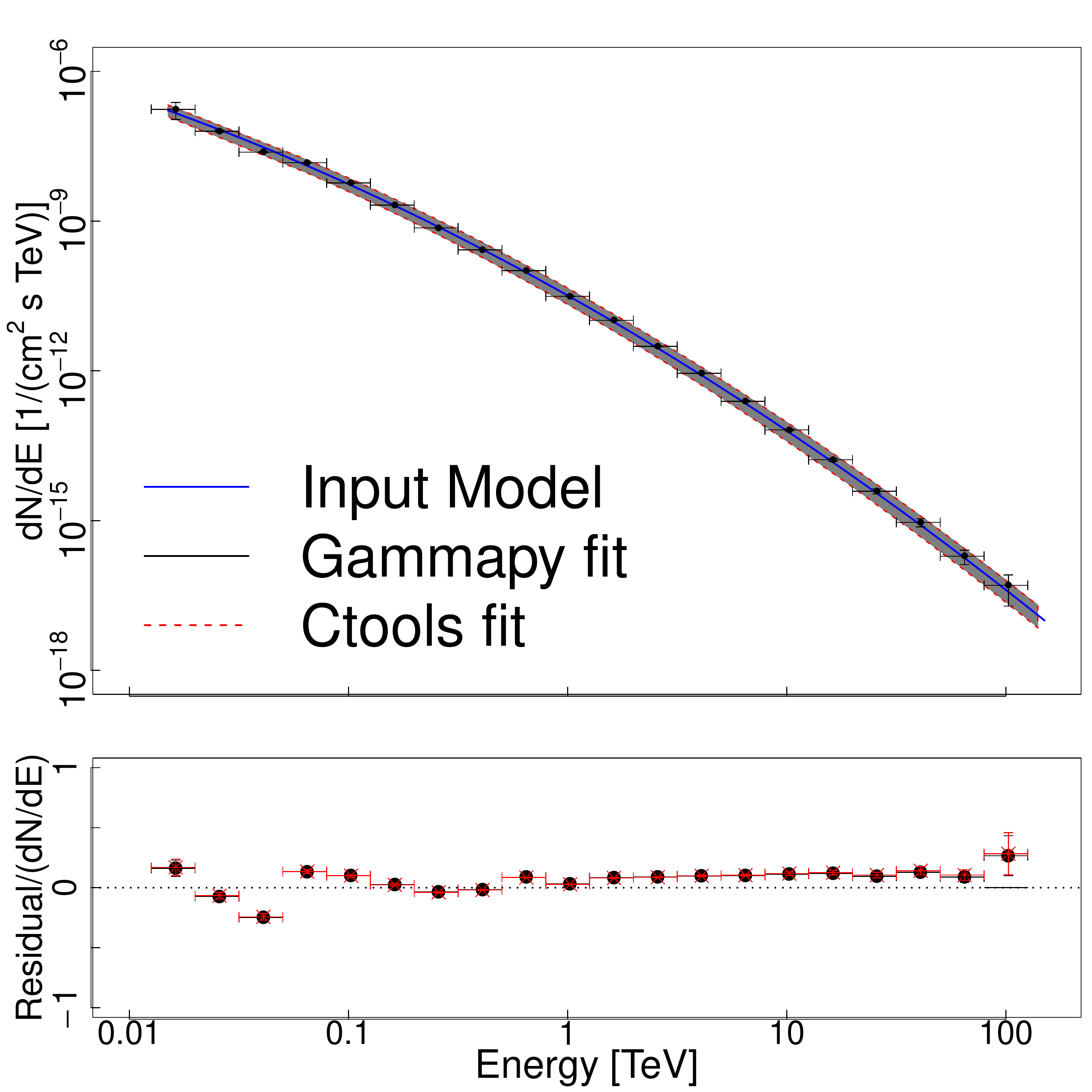}
\includegraphics[width=0.45\textwidth]{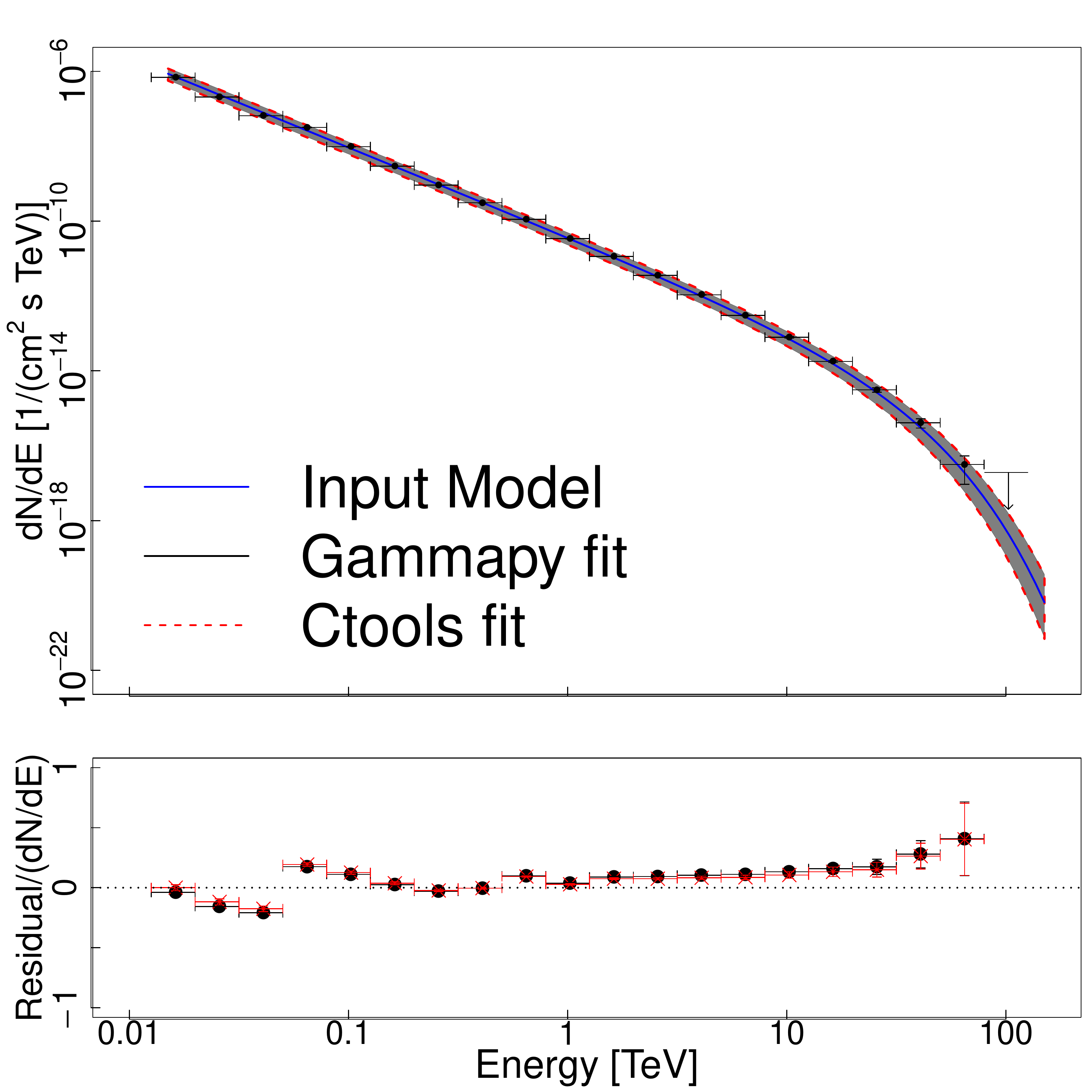}
\caption{Results of the simulations with LP-MAGIC (left) and a PLEC-HESS (right) models fitted with {\it Gammapy} and {\it ctools}.}
\label{fig:Gammapyctoolscomparison}
\end{figure*}

Figure \ref{fig:Gammapyctoolscomparison} proves that the spectral fitting results of {\it Gammapy} and {\it ctools} are compatible between each other.
Also, {\it ctools} and {\it Gammapy} fitting results present similar statistical errors.

\end{document}